\documentclass[aps,prl,twocolumn,showpacs,superscriptaddress]{revtex4-1}
\usepackage{graphicx}
\usepackage{amsmath,amsfonts}
\usepackage[usenames]{color}

\newcommand{\tauQ}{\tau_{\rm Q}}

\begin{document}

\title{Double universality of a quantum phase transition in spinor condensates:\\
       the Kibble-\.Zurek mechanism and a conservation law
       }

\author{Tomasz \'Swis\l{}ocki}
\affiliation{Instytut Fizyki PAN, Aleja Lotnik\'ow 32/46, 02-668 Warsaw, Poland}

\author{Emilia Witkowska}
\affiliation{Instytut Fizyki PAN, Aleja Lotnik\'ow 32/46, 02-668 Warsaw, Poland}

\author{Jacek Dziarmaga}
\affiliation{Instytut Fizyki Uniwersytetu Jagiello\'nskiego, ul. Reymonta 4, 30-059 Krak\'ow, Poland}

\author{Micha\l{} Matuszewski}
\affiliation{Instytut Fizyki PAN, Aleja Lotnik\'ow 32/46, 02-668 Warsaw, Poland}

\begin{abstract}
We consider a phase transition from antiferromagnetic to phase separated ground state in a spin-1 Bose-Einstein
condensate of ultracold atoms. We demonstrate the occurrence of two scaling laws, for the number of spin
fluctuations just after the phase transition, and for the number of spin domains in the final, stable configuration.
Only the first scaling can be explained by the standard Kibble-\.Zurek mechanism. 
We explain the occurrence of two scaling laws by a model including post-selection of spin domains
due to the conservation of condensate magnetization.
\end{abstract}
\pacs{03.75.Kk, 03.75.Mn, 67.85.De, 67.85.Fg}

\maketitle

The idea of a nonequilibrium phase transition has been attracting great attention
in many branches of physics. The number of physical models where it
has been considered or observed is steadily growing, now including not only the dynamics of the early Universe~\cite{Kibble}
and superfluid Helium~\cite{Zurek,Helium_KZ}, but also superconductors~\cite{superconductors_KZ}, cold atomic gases~\cite{BEC_KZ}, 
and other systems~\cite{other_KZ}.
The most notable outcome of such a phase transition is the possible creation of defects, such as
monopoles, strings, vortices or solitons~\cite{Kibble,Zurek,Damski_SolitonCreation} after 
crossing the critical point at a finite rate.
The Kibble-\.Zurek mechanism (KZM) is a theory that allows to predict the density of created defects
from the knowledge of the correlation length $\hat{\xi}$ at the instant when the system goes out of equilibrium~\cite{Zurek}.
The resulting scaling displays universal behaviour and is dependent only on the critical exponents
of the system $\nu$ and $z$.

Quantum phase transition, in contrast to a classical (thermodynamic) one,
occurs when varying a physical parameter leads to a change of the nature of the ground state~\cite{QPT_Book}.
Recently, a few theoretical works demonstrated that the KZM can be successfully applied to describe quantum phase
transitions in several models~\cite{Zurek_QPT,Dziarmaga_QPT,Damski_LZ,Damski_Ferromagnetic,Davis}, 
see Ref. \cite{Polkovnikov_Nonequilibrium} for reviews. 
Among these, Bose-Einstein condensates of ultracold atoms
offer realistic models of highly controllable and tunable systems~\cite{Damski_Ferromagnetic,Davis}. In~\cite{Davis}, 
the miscibility-immiscibility phase transition leading to the formation of stable, stationary domains
was proposed as an ideal candidate to observe KZM scaling in a quantum phase transition.

In this Letter, we investigate the critical scaling of the number of defects created during the transition to the
phase separated state of an antiferromagnetic spin-1 condensate~\cite{Miesner,Matuszewski_PS,Matuszewski_AF}. 
The introduction of a weak magnetic field
can lead in these systems to the transition from an antiferromagnetic ground state to a state where domains of atoms
with different spin projections separate~\cite{Matuszewski_PS}. In contrast to the transition considered in~\cite{Davis},
our scheme does not require the use of microwave coupling field or Feshbach resonances, which makes the experiment
simpler and more stable against inelastic losses. 

By employing numerical simulations within the truncated Wigner approximation,
we make a quite unexpected observation. 
While the number of spin fluctuations just after the phase transition closely follows the predictions of the
KZM, the number of spin domains in the final, stabilized state is given by a scaling law with a different exponent. 
We explain this double universality and predict the value of the second exponent 
using a model of system dynamics including
later stages of evolution, no longer described by the standard KZM. We show that when the nonlinear
processes set in, an effective post-selection of spin domains takes place, after which only a part of them
can survive due to the conservation of the condensate magnetization.

We consider a dilute antiferromagnetic spin-1 BEC in a homogeneous magnetic field pointing  along the $z$ axis.
We start with the Hamiltonian $H = H_0 + H_{\rm A}$, where the symmetric (spin independent) part is
\begin{equation} \label{En}
H_0 = \sum_{j=-,0,+} \int d x \, \psi_j^\dagger \left(-\frac{\hbar^2}{2m}\nabla^{2} + \frac{c_0}{2} n 
+ V(x)\right) \psi_j, \nonumber
\end{equation}
where the subscripts $j=-,0,+$ denote sublevels with magnetic quantum numbers along the magnetic field axis $m_f=-1,0,+1$,
$m$ is the atomic mass, $n=\sum n_j = \sum \psi_j^\dagger \psi_j$ is the total atom density and 
$V(x)$ is the external potential. Here we restricted the model to one dimension, with the other degrees of freedom confined
by a strong transverse potential with frequency $\omega_\perp$. The spin-dependent part can be written as
\begin{equation} \label{EA}
H_{\rm A} = \int d x \, \left(\sum_j E_j n_j + \frac{c_2}{2} :{\bf F}^2:\right)\,, \nonumber
\end{equation}
where $E_j$ are the Zeeman energy levels and the spin density is 
${\bf F}=(\psi^{\dagger}F_x\psi,\psi^{\dagger}F_y\psi,\psi^{\dagger}F_z\psi)$
where $F_{x,y,z}$ are the spin-1 matrices and $\psi =(\psi_+,\psi_0,\psi_-)$.
The spin-independent and spin-dependent interaction coefficients are given by 
$c_0=2 \hbar \omega_\perp (2 a_2 + a_0)/3>0$ and $c_2= 2 \hbar \omega_\perp (a_2 -
a_0)/3>0$, where $a_S$ is the s-wave scattering length for colliding atoms
with total spin $S$.
The total number of atoms $N = \int n d x$ and magnetization 
$M = \int \left(n_+ - n_-\right) d x$ are conserved.

The linear part of the Zeeman effect induces a homogeneous rotation of the spin vector around the direction of the magnetic field.
Since the Hamiltonian is invariant with respect to such spin rotations, we 
consider only the effects of the quadratic Zeeman shift~\cite{Matuszewski_PS}.
For sufficiently weak magnetic field we can approximate it by a positive energy shift of the $m_f=\pm 1$ sublevels $\delta
=(E_+ + E_- - 2E_0)/2  \approx B^2 A$, where $B$ is the magnetic field strength and $A=(g_I + g_J)^2 \mu_B^2/16 E_{\rm HFS}$,
$g_I$ and $g_J$ are the gyromagnetic ratios of electron and nucleus, $\mu_B$ is the Bohr magneton, 
$E_{\rm HFS}$ is the hyperfine energy splitting at zero magnetic field \cite{Matuszewski_PS}.

\begin{figure}[t]
\includegraphics[width=8.5cm]{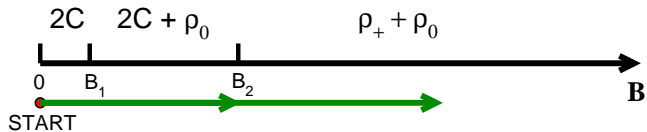}
\caption{Ground state phase diagram of an antiferromagnetic condensate for magnetization $M=N/2$.
We increase $B$ linearly during the time $\tauQ$ to drive the system through one or two phase transitions 
into a phase separated state. 
}
\label{scenario}
\end{figure}

The ground state phase diagram, shown in Fig.~\ref{scenario}(a), contains three phases divided by two critical points at
$B_1=B_0 M/\sqrt{2} N$ and $B_2=B_0/\sqrt{2}$, where $B_0=\sqrt{c_2 n/A}$ and $n$ is the total density. The ground state can be 
$(i)$ antiferromagnetic ($2C$) with $\psi=(\psi_+,0,\psi_-)$ for $B < B_1$,
$(ii)$ phase separated into two domains of the $2C$ and $\psi=(0,\psi_0,0)$ type ($\rho_0$)
for $B\in(B_1, B_2)$, or
$(iii)$ phase separated into two domains of the $\rho_0$ and $\psi=(\psi_+,0,0)$ type ($\rho_+$) for $B>B_2$~\cite{Matuszewski_PS}.
Note that the phase diagram is different in the special cases $M=0,\pm N$, which we disregard in the current paper.

For simplicity, we consider a system in the ring-shaped quasi-1D geometry with periodic boundary conditions at $\pm L/2$ and $V(x)=0$.
The magnetic field is initially switched off, and the atoms are prepared in the antiferromagnetic ($2C$) ground state
with magnetization fixed to $M=N/2$ (without loss of generality). To investigate KZM we increase $B$ linearly during the time $\tauQ$ to drive the system through one or two phase transitions into a phase separated state. Due to the finite quench time, the system ends up in a state with multiple spin domains.

The Kibble-\.Zurek theory is a powerful tool that allows to predict the density of topological defects resulting
from a nonequilibrium phase transition without solving the full dynamical equations.
The concept relies on the fact that the system does not follow the ground state exactly in the vicinity of the critical point due to the divergence of the relaxation time. 
The dynamics of the system cease to be adiabatic at 
$t\simeq-\hat t$
(here we choose $t=0$ in the first critical point),
when the relaxation time becomes comparable to the inverse quench rate
\begin{equation} \label{tau_freeze}
\hat{\tau}_{\rm rel} \approx |\hat{\varepsilon} / \hat{\dot{\varepsilon}}|,
\end{equation}
where 
$\varepsilon(t)=B - B_1\sim t/\tau_Q$
is the distance of 
the system from the critical point.
At this moment, the fluctuations approximately freeze, until the relaxation time
becomes short enough again. After crossing the critical point, distant parts of the system choose 
to break the 
symmetry in different ways, which
leads to the appearance of multiple defects in the form of 
domain walls between domains of 
$2C$ and $\rho_0$ phases. The average number of
domains is related to the correlation length 
$\hat\xi$ 
at the freeze out time 
$\hat{t} \sim \tauQ^{z\nu / (1+ z\nu)}$~\cite{Zurek,Polkovnikov_Nonequilibrium}
\begin{equation} \label{KZ_scaling}
N_{\rm d} = L / \hat{\xi} \sim \tauQ^{-\nu/(1+z\nu)},
\end{equation}
where $z$ and $\nu$ are the critical exponents determined by the scaling of the relaxation time 
$\tau_{\rm rel} \sim |\varepsilon|^{-z\nu}$ and excitation spectrum $\omega \sim |k|^z$, with $z=1$ in the superfluid.

\begin{figure*}
\includegraphics[width=18cm]{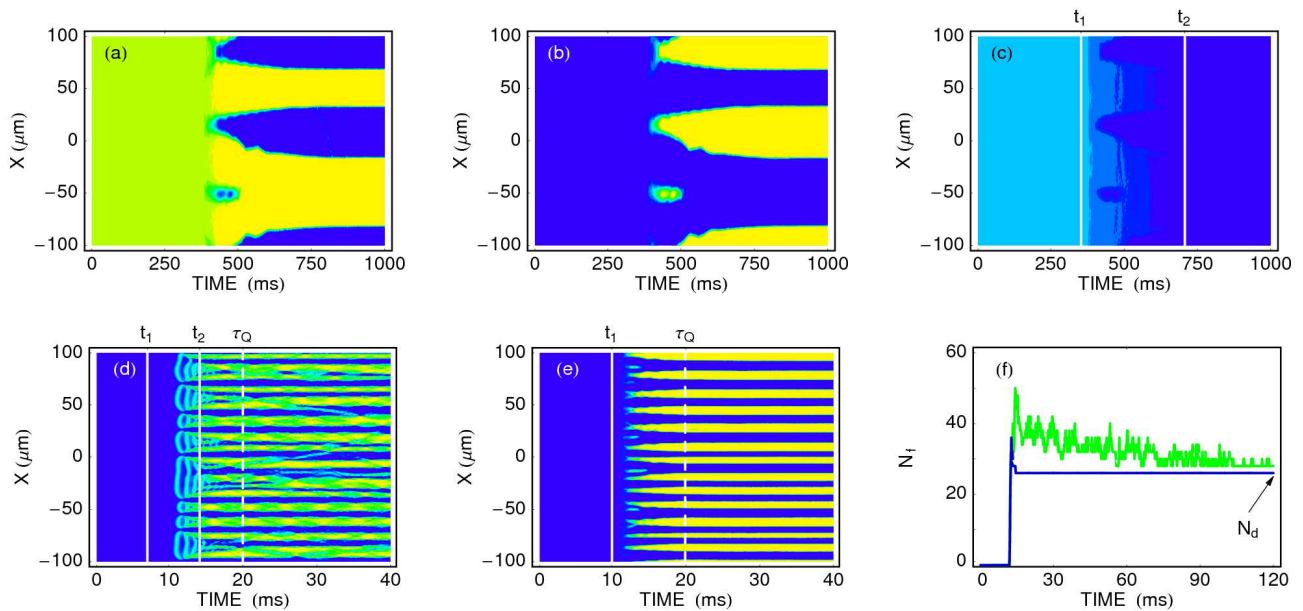}
\caption{
Spin domain formation dynamics in a ring-shaped 1D geometry with ring length $L=200\,\mu$m and $\omega_\perp=2\pi \times 1000\,$Hz. 
The top row shows densities (lighter is higher) of
$\psi_{+}$ (a), $\psi_{0}$ (b), and $\psi_{-}$ (c) for $N=10^{6}$ atoms, quench time $\tauQ=1\,$s, and final 
magnetic field strength $B/B_0=1$. The vertical lines at $t_1$ and $t_2$ correspond to the two phase transitions from Fig.~\ref{scenario}(a).
In (d) and (e), faster quenches with $\tauQ=20\,$ms are shown. The number of atoms in (e) 
is increased to $N=20\times 10^{6}$ and final
magnetic field is $(B/B_0)^2=0.49$ to show a cleaner process with a single phase transition and without interaction of fully formed domains. 
Frame (f) shows the number of spin fluctuations in function of time for the cases shown in (d), light line, and (e), dark line.
}
\label{dynamics}
\end{figure*}

We test the above prediction in numerical simulations within the truncated Wigner approximation,
with $N=10^6$ and parameters close to that of previous experiments in $^{23}$Na~\cite{Miesner}.
Additionally, we consider a "cleaner" case with transition through the first critical point only and  $N=20\times 10^{6}$ in order to minimize merging of domains.
Typical results of a single run, which can be interpreted as a single realization, are shown in Fig.~\ref{dynamics}.
We can clearly see the process of domain formation after the first phase transition at $t_1$.
However, 
there is always some number of spin fluctuations that disappear instead of evolving into full domains, see Fig.~\ref{dynamics}(f).

\begin{figure}
\includegraphics[width=8.5cm]{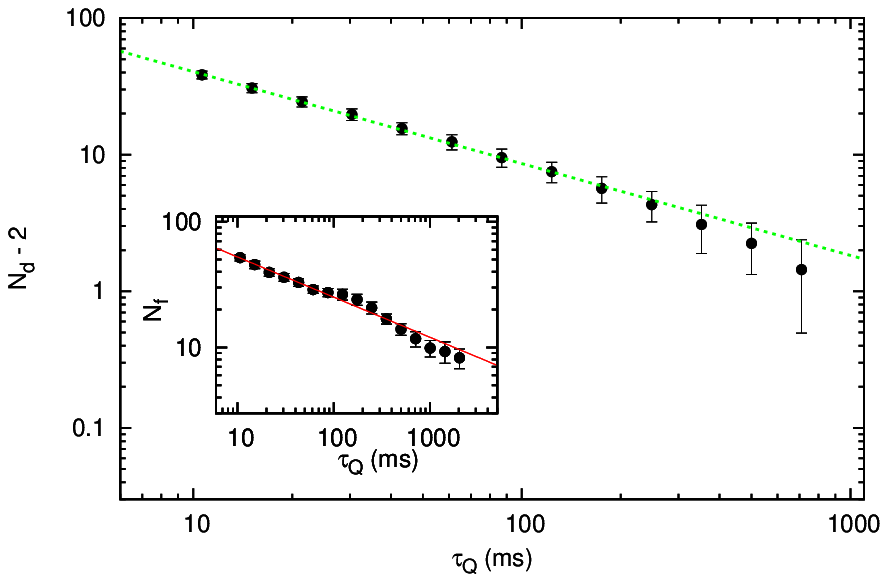}
\caption{Averaged number of spin domains after the quench as a function of the quench time for $N=20\times 10^{6}$. 
The scale is logarithmic on both axes
and $N_{\rm d}$ is decreased by two to account for the ground state phase separation into two domains. The points are
results of truncated Wigner simulations averaged over 100 runs. The dashed line is the fit to the power law with 
scaling exponent $n_{\rm d}=-0.67 \pm 0.01$. The inset shows
the maximal number of spin fluctuations counted just after the phase transition. The scaling exponent is here found to be 
$n_{\rm f}=-0.32 \pm 0.01$. 
}
\label{LL_20M}
\end{figure}

The above dynamics have a striking effect on the number of defects that are created in the system. 
In Fig.~\ref{LL_20M} we show the average number of defects in the function of the quench time $\tauQ$ for the ``cleaner'' case of 
large number of atoms, as in Fig.~\ref{dynamics}(e).
The critical exponents, calculated from the Bogoliubov excitation spectrum of the relevant gapped mode with
$\Delta_B=c_2n \sqrt{(\delta/c_2n - 1)^2-(1-M^2/N^2)}$ are identical as
in the case of a ferromagnetic~\cite{Damski_Ferromagnetic} or two component~\cite{Davis} condensate, $z=1$ and $\nu=1/2$. According to
the formula (\ref{KZ_scaling}), we could expect the scaling $N_{\rm d} \sim \tauQ^{-1/3}$. 
However, the number of domains in the final, stabilized state 
scales approximately as $N_{\rm d} \sim \tauQ^{-2/3}$ in a wide range of $\tauQ$, as depicted with the dashed line. 
Nevertheless, if we count the number of spin fluctuations~\cite{counting_fluctuations} just after the first phase transition, we do
recover the $N_{\rm f} \sim \tauQ^{-1/3}$ dependence (inset).

\begin{figure}
\includegraphics[width=8.5cm]{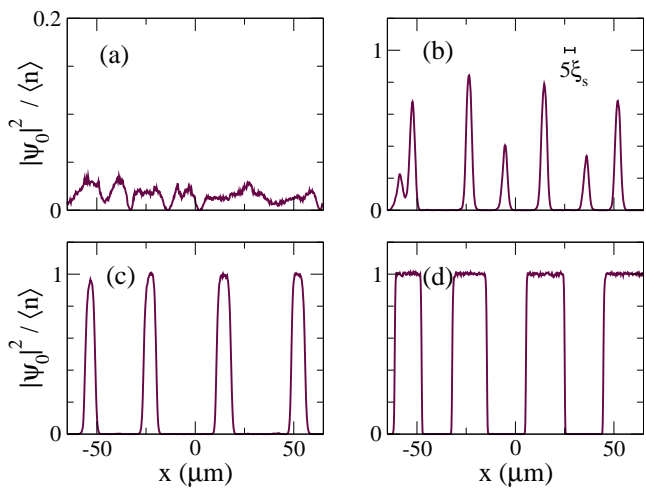}
\caption{
Profiles of the density $|\psi_0|^2$ for $\tau_{\rm Q}=100\,$ms, showing four consecutive
phases of the domain formation process after crossing the first critical point. 
(a) At $t=52.5\,$ms the small spin fluctuations begin to grow exponentially.
(b) At $t=56.25\,$ms the fluctuations transform into narrow (several $\xi_{\rm s}$) bubbles of $\psi_0$. 
(c) At $t=61.25\,$ms, as the bubbles mature, the post-selection eliminates some of them to keep the magnetization conserved. 
(d) At $t=100\,$ms, the domains have gradually increased in size and occupy half of the available area at $B=2B_1$.
The formation of the stable $\rho_0$-bubbles in the stages (a,b,c) takes place in the narrow time interval $t=52.5....61.25$ around $\hat t$. 
The magnetic field $\hat B$ within this relatively short time span satisfies $\hat B-B_1\sim\hat t/\tauQ\sim\tauQ^{-2/3}$.
}
\label{snapshots}
\end{figure}

We explain this puzzling appearance of two different scaling exponents with a model that includes
several phases of the domain formation process, see Fig.~\ref{snapshots}. 
The $2C$ state becomes unstable shortly after crossing the first critical point at $B=B_1$. Initially, the 
system follows the standard Kibble-\.Zurek scenario, as the spin fluctuations start to grow exponantially
at the freeze out time $\hat{t} \sim \tauQ^{1/3}$, see Fig.~\ref{snapshots}(a).
However, when the fluctuations become sufficiently large,
the nonlinearity sets in, which is visible as the formation of narrow bubbles of the $\psi_0$ component, see
Fig.~\ref{snapshots}(b).
This can be seen as the cooling of the system to the true ground state, which is now the phase separated $2C$ + $\rho_0$
state. As the bubbles of $\psi_0$ grow, they narrow to keep the magnetization approximately
conserved in their own neighborhood. We now count the number of small bubbles~\cite{counting_fluctuations},
which scales in the same way as the number of fluctuations in the Kibble-\.Zurek theory, $N_f \sim \tauQ^{-1/3}$.
At this point the KZ mechanism is completed, but it turns out to be just a preludium to 
the ultimate post-selection mechanism that sets the final density of domains $N_d$.

After a relatively short time of the growth, 
as compared to $\hat{t}$,
the density of $\psi_0$ in some of the bubbles manages to reach
its maximum value $n_0\approx n$, forming an array of $\rho_0$ domains. Their width is equal to several spin healing
lengths $\xi_s=(B_0/B)^2\hbar /\sqrt{2 m c_2 n}$, which is a natural scale defining the minimal width of a $\rho_0$ domain, 
see Fig.~\ref{snapshots}(c). 
However, it turns out that some of the bubbles must have disappeared to keep the magnetization conserved.
Indeed, it can be shown that in the $2C$ + $\rho_0$ ground state the fraction of the system occupied by the
$\rho_0$ phase is equal to $x_0 = 1 - B_1/B$. As we are still very close to the critical point, 
near $\hat B$ such that $\hat B-B_1\sim\hat t/\tauQ\sim\tauQ^{-2/3}$,
this fraction is small and 
scales with $\tauQ$ as 
$\hat{x}_0=1-B_1/\hat{B} \sim \tauQ^{-2/3}$. 
Since the spatial size of the bubbles is of the order of several $\xi_s$, independently of $\tauQ$, we now have $N_d \sim \tauQ^{-2/3}$ stable domains.
This is the final post-selected density of domain walls compatible with the conserved magnetization.

\begin{figure}
\includegraphics[width=8.5cm]{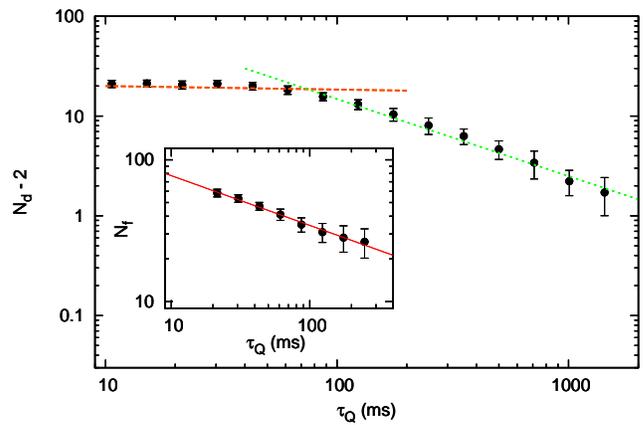}
\caption{Similar as in Fig.~\ref{LL_20M}, but for a realistic case of $N=10^{6}$. 
Here, the horizontal dashed line shows the saturation of the number of domains due to the finite spin healing length. 
The scaling exponents in this case
are $n_{\rm d}=-0.71 \pm 0.03$ and $n_{\rm f}=-0.35 \pm 0.01$.
}
\label{LL_1M}
\end{figure}

Once the domains are stabilized
they gradually grow in size with $x_0=1-B_1/B$ as the magnetic field is further increased, see Fig.~\ref{snapshots}(d).
In this last long stage of the evolution the domains are stable except for possible phase ordering kinetics~\cite{Bray}. 
This is a slow process where fluctuations eventually merge some domains slowly reducing the domain number $N_d$. 
The fluctuations are stronger for a lower number of atoms.
In Figure~\ref{LL_1M}, we plot the number of created defects for a lower (realistic) number of atoms $N=10^6$~\cite{Miesner}.
The interactions are much weaker in this case, which results in a larger spin healing length $\xi_{\rm s}$, 
which sets the smallest
possible domain size and is responsible for the saturation of the number of domains for 
small $\tau_Q$. Nevertheless, we can still clearly see  the two different 
scaling laws for $N_f$ and $N_d$ in a wide range of quench times $\tauQ$.
Here, the scaling exponents 
appear
to be slightly smaller than in the previous, ``cleaner'' case. 
Since the final state with multiple domains is not a true ground state, but a metastable state~\cite{Miesner}, 
we attribute this difference to the phase ordering kinetics, which effectively decrease 
the scaling exponents~\cite{Bray}.
However, the slow phase ordering taking place on a time scale much longer than $\hat t$ should be distinguished from 
the post-selection that happens near the same
$\hat t$ as the KZ mechanism.

In conclusion, we investigated a nonequilibrium phase transition in a relatively simple and stable experiment in
an antiferromagnetic Bose-Einstein condensate. 
We demonstrated the occurrence of two scaling laws describing the 
number of spin fluctuations and the final number of
spin domains. The occurrence of two scaling laws was explained in a model of system dynamics including an effective post-selection of 
spin domains due to the conservation of magnetization.
The post-selection transforming the scaling law is a general mechanism that should be effective whenever the standard Kibble-\.Zurek mechanism 
is not compatible with an additional conservation law. 

We thank Wojciech \.Zurek for reading the manuscript and useful comments.
This work was supported by the Polish Ministry of Science and Education grants 
N N202 128539 and IP 2011 034571, the National Science Center grant DEC-2011/01/B/ST3/00512,
and by the Foundation for Polish Science through the
\textquotedblleft Homing Plus\textquotedblright\ program.

\clearpage


\begin{thebibliography}{99}
\bibitem{counting_fluctuations} We count
the number $N_{\rm f}$ of zero crossings of the function $f(x)=n_0(x) - \alpha N/L$, where $\alpha=0.03$,
at the time instant when $N_{\rm f}$ is the largest. We checked that this method is accurate and weakly dependent 
on the choice of $\alpha$.
\end{thebibliography}
\end{document}